\documentclass[showpacs,amssymb,amsmath,aps,twocolumn,pra]{revtex4}
\newcommand{\half}{\mbox{$\frac{1}{2}$}}
\usepackage{graphics}
\usepackage{bm}
\begin{document}
\title{Global thermal entanglement in $n$-qubit systems}
\author{R. Rossignoli, N. Canosa}
\affiliation{Departamento de F\'{\i}sica, Universidad Nacional de La Plata,
C.C.67, La Plata (1900), Argentina}
\begin{abstract}
We examine the entanglement of thermal states of $n$ spins interacting through
different types of $XY$ couplings in the presence of a magnetic field, by
evaluating the negativities of all possible bipartite partitions of the whole
system and of subsystems. We consider both the case where every qubit interacts
with all others and where just nearest neighbors interact in a one-dimensional
chain. Limit temperatures for non-zero negativities are also evaluated and
compared with the mean field critical temperature. It is shown that limit
temperatures of global negativities are strictly independent of the magnetic
field in all $XXZ$ models, in spite of the quantum transitions that these
models may exhibit at zero temperature, while in anisotropic models they always
increase for sufficiently large fields. Results also show that these
temperatures are higher than those limiting pairwise entanglement.
\end{abstract}
 \maketitle
\section{Introduction}

Quantum entanglement \cite{S.35} is one of the most fundamental and intriguing
features of composite quantum systems, whose potential for developing radically
new forms of information transmission, processing and storage was only recently
recognized \cite{Be.93,Di.95,Be.00,NC.00}. Interest on the subject has
therefore grown considerably in recent years, but many aspects of entanglement,
particularly in mixed states of $n$-component systems, are still not fully
understood. Thermal entanglement \cite{N.98,Ar.01} refers normally to that of
mixed states of the form $\rho(T)\propto\exp[-H/T]$, with $T$ the temperature
and $H$ the system Hamiltonian, which are for instance the natural initial
states in NMR based quantum computing \cite{NC.00}.

A mixed state $\rho$ of a two component system $A+B$ is said to be entangled if
it cannot be written as $\sum_\alpha q_\alpha
\rho^\alpha_A\otimes\rho^\alpha_B$, with $q_\alpha>0$ and $\rho_{A,B}^\alpha$
density matrices for each component \cite{W.89}. If such an expansion is
feasible, $\rho$ is termed separable or classically correlated, since it is a
statistical mixture of product densities and the correlations between $A$ and
$B$ are then amenable to a classical description. A separable pure state
($\rho^2=\rho$) is always a product state $\rho_A\otimes\rho_B$, but this is
not necessarily the case for mixed states, where it is in general difficult to
prove separability. Moreover, in contrast with pure states \cite{BB.96}, there
is no unambiguous computable measure of the entanglement of mixed states,
except for two-qubit systems \cite{W.98}. Nonetheless, it is known that any
state of a $d$-dimensional system is separable if it is sufficiently close to
the fully mixed state $I_d/d$ \cite{ZHSL.98,B.99,GB.02} (i.e., if ${\rm
Tr}(\rho-I_d/d)^2\leq [d(d-1)]^{-1}$ in bipartite systems \cite{GB.02}). This
ensures the existence of a {\it finite limit temperature} for entanglement in
any finite interacting system, above which $\rho(T)$ becomes separable.

The situation is more complex in $n$-component systems \cite{DC.00}, where
there are first many possible bipartite splittings (bipartitions) of the whole
system to be considered. In addition, the separability of all bipartitions does
not warrant the representation of $\rho$ as a convex combination of $n$-product
densities $\otimes_{i=1}^n\rho_i$ (full separability), nor of $m$-product
($2<m<n$) densities ($m$-separability) \cite{DC.00}. There are as well many
$m$-component ($m<n$) subsystems whose reduced densities may similarly possess
different levels of entanglement. Given the lack of simple global measures,
many studies of interacting spin systems have then focused just on the
entanglement of the reduced pair density, which, though physically very
important, constitutes just a single aspect of the problem and leaves open the
question about the entanglement of the system as a whole.

The aim of this work is to study in more detail the thermal entanglement of
$n$-qubit systems by considering all possible bipartite splittings of the whole
system, as well as of selected subsystems, and evaluating the concomitant {\it
negativity} \cite{ZHSL.98,Z.99,HHH.00,VW.02}. This quantity is a measure of the
degree of violation of the Peres criterion for separability \cite{P.96,HHH.96},
and satisfies in addition some fundamental properties \cite{VW.02} which make
it a suitable measure of bipartite entanglement in mixed states. As physical
system we will consider $n$ spins interacting through $XYZ$ type couplings with
varying anisotropies, acting either between all spins or just between nearest
neighbors in a one-dimensional cyclic chain \cite{S.99,LSM.61}, and embedded in
a uniform magnetic field. These models are significant for solid-state based
qubit representations \cite{LDV.98,LDV.99,I.99,MSS.99,MSS.01,V.04} (the former
is relevant for schemes based on quantum dots electron spins \cite{I.99} and
Josephson junction arrays \cite{MSS.99,MSS.01,V.04}), and many relevant studies
of the two-qubit thermal entanglement in one-dimensional chains have been made
\cite{N.98,Ar.01,ON.02,W.01,GKVB.01,Wi.02,KS.02,GBF.03,SCC.03,CR.04}.

The picture that will here emerge is that of a hierarchy of negativities of
global and reduced bipartitions which will possess different limit
temperatures. The system will loose its quantum correlations as $T$ increases
through a cascade of ``transitions'' that indicate the onset of separability of
the different bipartitions, with reduced pair densities becoming separable
before global partitions. The behavior of all negativities depends strongly on
the interaction. In $XXZ$ models, limit temperatures of global negativities are
remarkably {\it independent} of the magnetic field, as will be demonstrated,
even though the ground state may exhibit quantum phase transitions as the field
is varied, while in anisotropic $XYZ$ models they always increase as the field
increases, even though ground state entanglement decreases. This behavior
confirms that observed in two-qubit systems \cite{KS.02,CR.04}. Comparison with
the mean field critical temperature will also be made, and shows that
symmetry-breaking mean field solutions are not necessarily indicators of
entanglement for $T>0$.

\section{Formalism}
We consider $n$ qubits or spins coupled through an $XYZ$ type interaction in
the presence of a uniform magnetic field along the $z$ axis. The Hamiltonian is
\begin{subequations}
\label{H}
\begin{eqnarray}
H&=&bS_z-\sum_{i<j}(v^{ij}_x s_x^is_x^j+v^{ij}_ys_y^is_y^j+v^{ij}_zs_z^is_z^j)
\label{H1}\\
&=&H_z-\sum_{i<j}(v^{ij}_+s_+^is_-^j+v^{ij}_-s_+^is_+^j+h.c.)\,,\label{H2}
\end{eqnarray}
\end{subequations}
where $\bm{s}^i$ are the spin operators, $S_z=\sum_{i=1}^ns_z^i$ is the total
spin $z$-component, $s^i_{\pm}=s^i_x\pm s^i_y$,
$v^{ij}_{\pm}=\frac{1}{4}(v^{ij}_x\pm v^{ij}_y)$ and
$H_z=bS_z-\sum_{i<j}v^{ij}_zs^i_zs^j_z$. We will consider two types of
interaction range: I) that where every spin interacts identically with all
others ($v^{ij}_\alpha=v_\alpha$ $\forall$ $i<j$ and $\alpha=x,y,z$) and II)
that where only nearest-neighbors interact within a 1D cyclic chain
($v_\alpha^{ij}=(\delta_{j,i+1}+\delta_{i1}\delta_{jn})v_\alpha$ for $i<j$ and
$\alpha=x,y,z$). Both types become coincident for $n=3$. Regardless of the
interaction range, $H$ always commutes with the $z$-parity or phase flip
$e^{i\pi S_z}$, and will commute as well with $S_z$ when $v^{ij}_x=v^{ij}_y$
$\forall$ $i,j$ ($XXZ$ models). The spectrum of $H$ is obviously independent of
the sign of $b$ and $v_-$ (in I and II).

For type I, $H$ is in addition invariant against any permutation of its qubits
and can be rewritten in terms of the total spin components
$S_\alpha=\sum_{i=1}^ns^i_\alpha$ as
\begin{equation}
H_I=bS_z-\half(v_xS_x^2+v_yS_y^2+v_zS_z^2)+E_0\,,\label{HI}
\end{equation}
with $E_0=n(v_x+v_y+v_z)/8$. It commutes therefore  with
$S^2=\bm{S}\cdot\bm{S}$. Its eigenvalues $E_{SL}$ can then be obtained by
diagonalizing $H$ in each representation with total spin $S\leq n/2$, of
dimension $2S+1$ and multiplicity $Y(S)=(^n_k)-(^{\;\;n}_{k-1})$, with $k=\half
n-S$ and $Y(n/2)=1$, such that $\sum_S(2S+1)Y(S)=2^n$. The eigenstates will be
of the form $|SL\alpha\rangle$, with $L=1,\ldots,2S+1$, and
$\alpha=1,\ldots,Y(S)$ a degeneracy index labelling different permutations.
Effective pseudospin Hamiltonians of this form have been much employed in
nuclear physics \cite{LMG.65,RS.80}, and are also suitable for describing the
effective interaction of Josephson junction based charge qubits \cite{MSS.01}
as well as of quantum dots electron spins coupled through microcavity modes
\cite{I.99}.

For type II, $H$ is translationally invariant, and for $v_z=0$ it can be
rewritten exactly for each $z$-parity as a quadratic fermionic form
\cite{LSM.61}. All $2^n$ eigenvalues of $H$ can then be obtained from the
ensuing quasiparticle energies, determined through a fermionic Bogoliubov
transformation. For even $n$, the spectrum of $H$ is in this case also
independent of the sign of $v_+$, since it can be changed just by inverting the
$x$ and $y$ directions at odd (or even) sites.

We shall examine the entanglement of the corresponding $n$-qubit thermal mixed
state
 \begin{equation}
 \rho=\exp[-H/T]/{\rm Tr}\exp[-H/T]\,,\label{1}
 \end{equation}
by considering {\it all} possible bipartitions of $k$ and $n-k$ qubits and
determining the associated {\it negativities} \cite{VW.02}, defined as the
absolute value of the sum of the {\it negative} eigenvalues of the ensuing
partial transpose $\rho^{t_p}$ of $\rho$:
\begin{equation}
N(\rho)=\half{\rm Tr}(|\rho^{t_p}|-\rho^{t_p})\,.
\end{equation}
 Here $\rho^{t_p}_{ij,kl}=\rho_{il,kj}$, with $i,k$ $(j,l)$ labels for
states of the first (second) component of the partition and $|\rho^{t_p}|=
\sqrt{(\rho^{t_p})^2}$. According to the Peres criterion \cite{P.96}, if
$N(\rho)>0$ the two components of the partition are entangled. Moreover,
$N(\rho)$ satisfies some fundamental properties of an entanglement measure
\cite{VW.02}: It does not increase under local operations and classical
communication (LOCC), being then an entanglement monotone, and is a convex
function of $\rho$ ($N(\sum_\alpha p_\alpha\rho_\alpha)\leq \sum_\alpha
p_\alpha N(\rho_\alpha)$ for $p_\alpha\geq 0$, $\sum_\alpha p_\alpha=1$). It
also provides an upper bound to the teleportation capacity and distillation
rate \cite{VW.02,HHH.00}. Not all aspects of entanglement can be captured in
this way, as Peres criterion is in general sufficient only for two-qubit or
qubit+qutrit systems (although entangled states satisfying $N(\rho)=0$ are
bound entangled), and the separability of all bipartitions does not imply full
separability \cite{DC.00}. We shall not examine these features but rather focus
on $N(\rho)$ as an indicator of useful bipartite entanglement. In the same way
we will examine the entanglement of reduced densities $\rho_m={\rm
Tr}_{n-m}\rho$ for $m<n$ selected qubits. It is apparent that for $b\neq 0$,
entanglement in the state (\ref{1}) can only be generated by the $XY$ terms in
(\ref{H1}), i.e., the sum in (\ref{H2}), as $H_z$ is diagonal in the basis of
separable eigenstates of $S_z$ (standard basis).

For interactions of type I, the negativities of bipartitions with $k$ and $n-k$
qubits will depend just on $n$ and $k$, since $\rho$ is here completely
symmetric under arbitrary permutations and any choice of the $k$ states is
equivalent. There are thus $[n/2]$ negativities $N^n_k$ ($=N^n_{n-k}$) that
characterize the global bipartite entanglement (i.e., $N^3_1$ for $n=3$,
$N^4_1$ and $N^4_2$ for $n=4$).

Reduced densities $\rho_m$ for $m<n$ qubits in I will also depend just on $m$,
as any choice of the $m$ states is equivalent, and will be as well completely
symmetric. We have then $[m/2]$ negativities $N^m_k$ of reduced bipartitions
with $m$ and $m-k$ qubits. Note that all negativities of reduced densities can
be zero even if $N^n_k>0$ $\forall$ $k$, as occurs for the well-known GHZ type
pure states ($|0\ldots 0\rangle+|1\ldots 1\rangle)/\sqrt{2}$, which lead to
$N^n_k=1/2$ and $N^m_k=0$ for $m<n$ $\forall$ $k$. On the other hand, if
$N^n_k=0$ $\forall k$, then $N^m_k=0$ $\forall$ $k$ and $m<n$. Actually, since
tracing out a part of a local subsystem is a LOCC operation \cite{VW.02}, we
have the inequality
 \begin{equation}
 N^{n-j-l}_{k-j}(\rho)\leq N^n_k(\rho)\,,\label{Nkr}
 \end{equation}
for $j<k$, $l<n-k$, as the bipartition with $k-j$ and $n-k-l$ qubits of
$\rho_{n-j-l}$ can be obtained by tracing out $j$ ($l$) qubits from the first
(second) component of a global partition with $k$ and $n-k$ qubits. This
implies $N^{n-j-l}_{k-j}=0$ if $N^n_k=0$ and hence the ordering of limit
temperatures $T^{n-j-l}_{k-j}\leq T^n_k$ in the same system. For example,
$N^2_1\leq N^3_1\leq N^4_k$ for $k=1,2$.

For type II, the negativities will depend as well on the spacings between the
$j$ states of a subsystem. For instance, for $n=4$ ordered qubits $abcd$ we
have three global negativities: $N^4_1\equiv N_{a-bcd}=N^4_3$, and $N^4_2\equiv
N_{ab-cd}$, $N^4_{2'}\equiv N_{ac-bd}$, corresponding to partitions with two
adjacent and two non-adjacent qubits in each subsystem respectively. Reduced
densities for $m$ qubits will depend as well on the spacings between the $m$
states, and will not be necessarily cyclic, so that all possible partitions
will have to be examined. For instance, for $m=3$ adjacent qubits $abc$ in a
cyclic chain of $n>3$ qubits, we have the adjacent $N^3_1\equiv
N_{a-bc}=N_{ab-c}$ and the non-adjacent $N^3_{1'}\equiv N_{b-ac}$ reduced
negativities. Thus, for $n=4$ we have 3 global negativities, a single type of
three-qubit reduced density $\rho_3$ with two negativities, and 2 two-qubit
densities $\rho_2$, $\rho_{2'}$ for adjacent ($ab$) and non-adjacent ($ac$)
qubits respectively, with negativities $N^2_1$, $N^{2'}_1$. By similar
arguments as above, for $n=4$ we obtain the hierarchies $N^3_{1,1'}\leq N^4_1$,
$N^3_{1}\leq N^4_2$, $N^3_{1'}\leq N^4_{2'}$, and $N^{2,2'}_1\leq N^3_1$,
$N^2_1\leq N^3_{1'}$.

{\it Independence of limit temperatures from the magnetic field when
$[H,S_z]=0$.} A remarkable feature of the thermal states (\ref{1}) with the
Hamiltonian (\ref{H}) is that when $v^{ij}_{-}=0$ $\forall i,j$ ($XXZ$ models),
the limit temperatures of {\it global} negativities are {\it independent} of
the applied magnetic field $b$, even though the negativities and the ground
state entanglement are not, since in such a case the eigenstates of $H$ do not
depend on $b$ and the field dependence of $\rho$ can be factorized. Writing the
states in the standard basis succinctly as $|m,m'\rangle$, where $m$ ($m'$)
denotes the $z$-component of spin in the first (second) subsystem of a
bipartition (remaining labels omitted), $\rho^{t_p}$ will have non-zero matrix
elements just between states $|m,m'\rangle$ and $|m+k,m'+k\rangle$, arising
from those of $\rho$ between states $|m,m'+k\rangle$ and $|m+k,m'\rangle$ with
the same total spin $M=m+m'+k$. The field dependence of these elements is
contained in a factor $e^{-bM/T}/Z(b)$, where $Z(b)={\rm Tr}e^{-H/T}$ is the
partition function. Hence, we may write
\begin{equation}
\rho^{t_p}(b)=r(b)\exp[-bS_z/2T]\rho^{t_p}(0)\exp[-bS_z/2T]\,,\label{rtj}
\end{equation}
with $r(b)=Z(0)/Z(b)\neq 0$ and $[\rho^{t_p}(0),S_z]\neq 0$. While the field
dependence of the eigenvalues of $\rho^{t_p}(b)$ cannot always be factorized,
we obtain, as ${\rm Tr}S_z=0$,
\begin{equation}
{\rm Det}\,\rho^{t_p}(b)=[r(b)]^{2^n}{\rm Det}\,\rho^{t_p}(0)\,,
\end{equation}
implying that the condition ${\rm Det}[\rho^{t_p}(b)]=0$, which determines in
particular the limit temperature for nonzero negativity (i.e., zero lowest
eigenvalue) is the same as that for $b=0$. Moreover, it is apparent from the
strict positivity of $e^{-bS_z/2T}$ for $T>0$ that $\rho^{t_p}(b)$ is positive
(no negative eigenvalues) if and only if $\rho^{t_p}(0)$ is positive. A similar
result for limit temperatures of negativities of reduced densities does not
hold.

This result has some interesting consequences. In particular, for $[H,S_z]=0$
the ground state will always become separable for sufficiently large fields
(i.e., the state with all spins aligned $|0\rangle\equiv |SM\rangle$, with
$S=|M|=n/2$), but the limit temperature will be the same for all fields,
implying the reentry of entanglement for $T>0$ for such values if $N(\rho)>0$
at low fields. It also implies that global limit temperatures will never
coincide with the mean field critical temperature since the latter depends on
the magnetic field (see below).

{\it Negativity for large fields when $[H,S_z]\neq 0$}. Another important
feature is that for sufficiently large $|b|$, the thermal density (\ref{1})
will possess at least one nonzero negativity at any finite temperature $T\ll
|b|$ if $v^{ij}_{-}\neq 0$ at least for some pair, implying a {\it divergence}
of the corresponding limit temperature for $|b|\rightarrow\infty$. For
sufficiently large fields, we may treat all interaction terms perturbatively.
If $|b|\gg T$, the thermal density will then approach that of the ground state
$|\phi_0\rangle$, which will be the aligned state $|0\rangle$ plus a small
perturbation, as the weight of excited states become exponentially small (of
order $e^{-|b|/T}$ or less). Assuming $b>0$, up to first order in
$v_{\alpha}^{ij}/b$ we have
$|\phi_0\rangle=[I+\sum_{i<j}(v^{ij}_-/2b)s_+^{i}s_+^j]|0\rangle$, which is an
entangled state. The negativity of a partition with $i$, $j$ in different
subsystems and $v^{ij}_-\neq 0$ will then be non-zero and of order $v_-/b$ in
this limit, with $v_-$ the order of the $v^{ij}_-$. For instance, in the fully
connected case I we obtain, for a partition with $k$ and $n-k$ qubits and $T\ll
|b|$,
\begin{equation}N^n_k\approx\half\sqrt{k(n-k)}|v_-/b|\label{nk1}\end{equation}
up to first order in $v_-/b$. Note that $k(n-k)$ is just the number of links
$v^{ij}_-$ between the two subsystems, being $N^n_k$ maximum for $k=[n/2]$. In
type II, for partitions with $k$ and $n-k$ {\it adjacent} qubits, we obtain
instead
\begin{equation}N^n_k\approx \half r_k|v_-/b|\,,\label{nk2}\end{equation}
where $r_1=\sqrt{2}$ and $r_k=2$ for $2\leq k\leq [n/2]$. For partitions
consisting of non-contiguous qubits the value of $r_k$ will be larger (and in
fact proportional to $n/2$ for a bipartition with $k=[n/2]$ non-adjacent qubits
in each set).

{\it Mean field approximation.} The thermal state (\ref{1}) provides the
absolute minimum of the free energy $F(\rho)=\langle H\rangle-TS(\rho)$, where
$\langle H\rangle={\rm Tr}\rho H$ and $S(\rho)=-{\rm Tr}\rho\ln\rho$ is the von
Neumann entropy. The f  inite temperature mean field approximation is based on
the minimization of $F(\rho)$ within the subset of uncorrelated density
operators
 \begin{equation}
 \rho_{\rm mf}(T)=\exp[-h/T]/{\rm Tr}\exp[-h/T],\;\;h=
 \sum_{i=1}^n\bm{\lambda}_i\cdot \bm{s}^i\,.
 \end{equation}
This leads to the self-consistent equations $\bm{\lambda}_i=\partial \langle
H\rangle_{\rm mf}/\partial \langle \bm{s}^i\rangle_{\rm mf}\,,$ with $\langle
O\rangle_{\rm mf}={\rm Tr}\rho_{\rm mf}O$, which determine a $T$ dependent (and
possibly symmetry-breaking) effective non-interacting hamiltonian $h$. For
$v_z=0$ and $v_x\geq 0$, with $|v_y|\leq v_x$, it suffices to consider
$h=bS_z-\lambda S_x$. The limit temperature of the ensuing $z$-parity symmetry
breaking solution ($\lambda\neq 0$) is
\begin{equation}
T_c=|b|/\ln[\frac{\eta+1}{\eta-1}],\;\;\;\;\eta\equiv v/|b|>1\,,
 \end{equation}
where $v=v_x(n-1)/2$ in I and $v=v_x$ in II. For fixed $v$, the mean field
solution and intensive energy
\begin{equation}
\langle H\rangle_{\rm mf}/n=b\langle s_z\rangle_{\rm mf}-
v\langle s_x\rangle^2_{\rm mf}\label{leh}
\end{equation}
are hence the same in I and II, and are independent of the number of qubits $n$
and of $v_y$.

For $\eta\gg 1$ (low fields $|b|\ll v$), $T_c/v=\half+O(\eta^2)$, while for
$\eta\leq 1$ ($|b|\geq v$), $T_c=0$ (i.e., $\lambda=0$ $\forall$ $T$) and no
interaction is ``seen'' at the mean field level. This implies that at least for
$|b|>v$, there will be no agreement between $T_c$ and the limit temperatures
for global negativities even for {\it large} $n$, as the latter either stay
constant ($XXZ$ models) or tend to increase (full anisotropic models) for
increasing fields. Note also from Eq.\ (\ref{nk1}) that for large fields
($\eta\ll 1$), the negativity of symmetric global partitions $(k=n/2)$ in
anisotropic models of type I remains non-zero for $n\rightarrow\infty$ even if
$v_-\propto v/n$, and the same occurs in type II models (for $v_-\propto v$),
explaining thus the lack of agreement with $T_c$. Nevertheless, for $\eta>1$
limit temperatures for global negativities  will typically be of the same order
as $T_c$, as will be seen in the next section.

Let us note as well that in the fully connected case I, the strength of the
common pair coupling $v^{ij}=2v/(n-1)$ between spins $s^i$,$s^j$ required for
symmetry breaking at finite temperature or field scales as $n^{-1}$, becoming
then smaller than in an array of type II as $n$ increases. Thus, at the mean
field level weak pair couplings in I have the same effect as much stronger
values in II. Typical values of $v^{ij}$ can be of order
$4\tilde{g}_{ij}\approx 0.08$ meV for quantum dots electron spins coupled
through a microcavity mode \cite {I.99}, while in Josephson charge qubits
arrays \cite{MSS.99}, $v_{ij}=4E^i_JE^j_J/E_L\approx 40mK$ (in temperature
units) for $E_L=10 E_J$, where $E^i_J\propto E_J\approx 100 mK$ are the
effective Josephson energies of the qubits controlled by the external fluxes
and $E_L$ an energy scale depending on the SQUID inductance $L$. For
$|b|\approx E_J$ we have then $\eta\approx 0.2(n-1)$ and $T_c\approx \half
v\approx 10 mK (n-1)$ for $\eta\agt 2$.

\section {Results}
\begin{figure}[t]
\vspace*{-1.5cm}

     \centerline{\scalebox{0.6}{\includegraphics{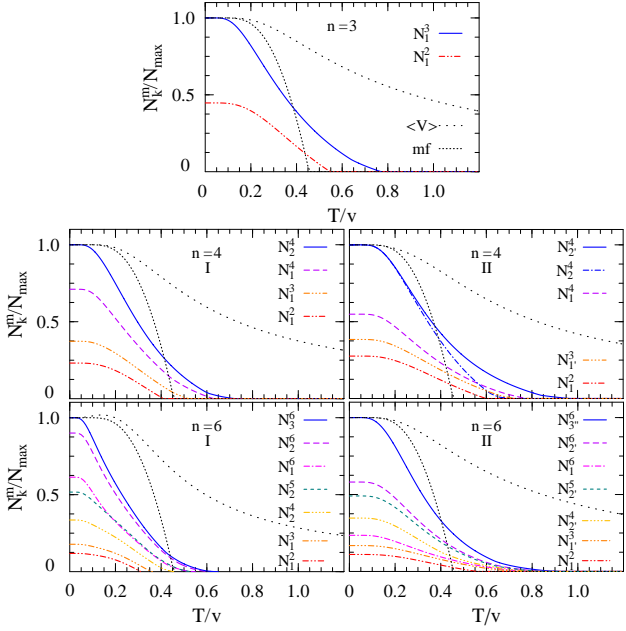}}}
    \vspace*{-8.cm}

\caption{Thermal behavior of the negativities determined by the state (\ref{1})
for $n=3$, $4$ and 6 qubits, for an $XY$ interaction with $v_y=-v_x$ ($v_+=0$)
and $v/b=2$. Results for both the full (I) and cyclic nearest neighbor (II)
interaction ranges, coincident for $n=3$, are depicted. $N^n_k$ is the
negativity of a global bipartition with $k$ and $n-k$ qubits, while for $m<n$,
$N^m_k$ is that of a reduced density for $m$ qubits, adjacent in II. $N_{\rm
max}$ is the largest negativity at $T=0$ and primes in panels II denote
different bipartitions (see text). Also shown for reference are the exact
thermal average of the interaction ($\langle V\rangle$) and its mean field
average (mf), both scaled for clarity to their $T=0$ values.} \label{f1}
\end{figure}

We discuss now numerical results for the $XY$ case ($v_z=0$) with $v_x\geq 0$
and different anisotropies $v_-/v_+=(v_x-v_y)/(v_x+v_y)$, for interactions of
types I and II. We first consider the thermal behavior in the fully anisotropic
case $v_y=-v_x$ ($v_+=0$), because it is the simplest to describe and
represents that of a system with a non-degenerate entangled ground state well
separated from the remaining states, which depends smoothly on the magnetic
field. Typical results for $n=3,4,6$ and $\eta=v/|b|=2$ are shown in fig.\
\ref{f1}. For $n=3$, there is a single global negativity $N^3_1=N^3_2$, which
decreases monotonously as $T$ increases, vanishing at $T^3_1\approx 0.77 v$,
while the negativity $N^2_1$ of the reduced two-qubit density is smaller and
vanishes at $T^2_1\approx 0.54v$. Hence, there is an appreciable interval
$[T_1^2,T^3_1)$ where only global entanglement persists. Both $T^3_1$ and
$T^2_1$ are {\it higher} than the mean field critical temperature $T_c\approx
0.46 v$, although of the same order, and negativities are actually rather small
above $T_c$. The mean value of the interaction $\langle V\rangle\equiv {\rm
Tr}\rho V$, where $V$ denotes the sum in Eq.\ (\ref{H}), remains however quite
significant for $T>T^3_1$, vanishing only for $T\rightarrow\infty$, which
indicates that $\rho$ remains considerably correlated for $T>T^3_1$ albeit in
an essentially classical manner.

The behavior for $n=4$ and $6$ qubits is similar. For type I interaction (left
panels) and $n=4$, there are just two global negativities, $N^4_1$, $N^4_2$,
the latter being the strongest and most persistent ($(T^4_1,T^4_2)\approx
(0.62,0.72)v$). This difference can be attributed to the higher number of links
$v^{ij}$ between both subsystems existing in the latter ($4$ for $N^4_2$  and
$3$ for $N^4_1$). The negativities $N^3_1$ and $N^2_1$ of the reduced three-
and two-qubit densities are smaller, in agreement with (\ref{Nkr}), and vanish
at $(T^3_1,T^2_1)\approx (0.52,0.41)v$ (which are lower than those of the upper
panel). There is again an interval $[T^3_1,T^4_2)$ where only {\it global}
negativities $N^4_1$, $N^4_2$ are non zero, and a smaller interval
$[T^4_1,T^4_2)$ where just {\it one} global negativity survives. For $n=6$ we
obtain similarly a cascade of limit temperatures $(T^6_1,T^6_2,T^6_3)\approx
(0.54,0.6,0.64)v$ for the global negativities $N^6_1$, $N^6_2$, $N^6_3$, the
latter being the greatest and most persistent in agreement with the number of
links (5, 8 and 9 respectively). There is as well a series of lower limit
temperatures
$(T^2_1,T^3_1,T^4_1,T^4_2,T^5_1,T^5_2)\approx(0.31,0.36,0.41,0.45,0.47,0.53)v$
for the negativities of reduced densities $\rho_m$. Just the most persistent
one for each $m$ ($N^m_{[m/2]})$ is shown. Note that $N^5_2\approx N^6_1$ for
$T\agt 0.2 v$, as $N^5_2$ is not necessarily smaller than $N^6_1$.

The behavior for type II interaction is similar, although limit temperatures
are higher (in comparison with the corresponding value of $v$ or $T_c$). For
$n=4$ qubits $abcd$, there are three different global negativities, the most
persistent being that of the {\it non-adjacent} 2+2 partition
$N^4_{2'}=N_{ac-bd}$, followed by $N^4_1=N_{a-bcd}$ and finally that of the
adjacent partition $N^4_2=N_{ab-cd}$. The latter, though coincident with the
first one at $T=0$, decays faster: $(T^4_{2},T^4_1,T^4_{2'})\approx
(0.71,0.78,0.96)v$. This is again in agreement with the number of links between
subsystems (4 for $N^4_{2'}$, $2$ for $N^4_2$ and $N^4_1$). The reduced pair
density of adjacent qubits $ab$ remains now entangled until $T^2_1\approx 0.59
v$, while that of two non-adjacent qubits $ac$ is here {\it separable}
$\forall$ $T$ ($N^{2'}_{1}=0$). The reduced three-qubit density has here two
negativities: $N^3_1=N_{a-bc}$ and $N^3_{1'}=N_{ac-b}$, the latter being the
most persistent: $(T^3_{1},T^3_{1'})\approx (0.62,0.72)v$. Just $N^3_{1'}$ is
shown. We have then the ordering
$T^2_1<T^3_{1}<T^4_{2}<T^3_{1'}<T^4_1<T^4_{2'}$. Note that $N^4_2$ is not
necessarily greater than $N^3_{1'}$.

For $n=6$ qubits $abcdef$ there are 7 global negativities for type II: $N^6_1$,
$N^6_2=N_{ab-cdef}$, $N^6_{2'}=N_{ac-bdef}$, $N^6_{2''}=N_{ad-bcef}$, and
$N^6_3=N_{abc-def}$, $N^6_{3'}=N_{abd-cef}$, $N^6_{3''}=N_{ace-bdf}$. The most
persistent is that of the non-contiguous 3+3 partition $N^6_{3''}$ (6 links),
followed by $N^6_{2'}$ (4 links), with
$(T^6_{3''},T^6_{2'})\approx(0.95,0.87)v$, whereas those of adjacent qubits,
$N^6_1$, $N^6_2$ and $N^6_3$ (2 links), are the first to vanish:
$(T^6_1,T^6_2,T^6_3)\approx(0.77,0.71,0.7)v$. Only the most persistent of each
$N^6_k$ set is depicted. Also shown are the most persistent negativities of
reduced densities of adjacent qubits, which are again those of most symmetric
partitions with non-adjacent qubits: $N^3_{1'}$ for $\rho_3$ and
$N^4_{2'}=N_{ac-bd}$, $N^5_{2'}=N_{bd-ace}$ for $\rho_4$, $\rho_5$ (which have
5 and 9 different negativities). We obtain the ordering
$T^2_1<T^3_{1'}<T^6_1<T^4_{2'}<T^5_{2'}<T^6_{2'}<T^6_{3''}$.

\begin{figure}[t]
\vspace*{-1.5cm}

 \centerline{\scalebox{0.6}{\includegraphics{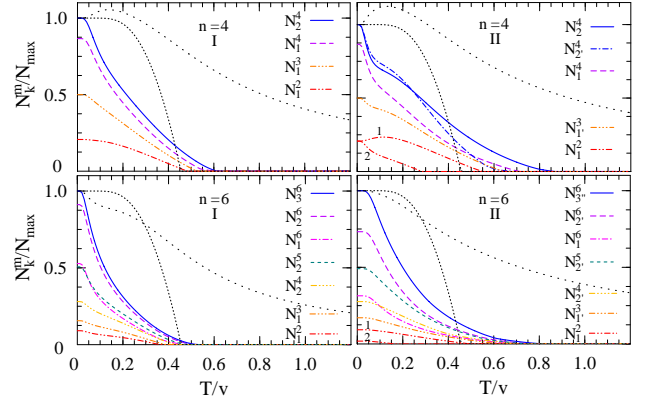}}}
    \vspace*{-11.5cm}

\caption{Same quantities as fig.\ \ref{f1} for $v_y=v_x$ ($v_-=0$). Details are
the same as before. Right panels 
 results for the reduced density of
both adjacent (1) and first non-adjacent (2) pairs. The latter vanished in
fig.\ 1.}\label{f2}
\end{figure}

\begin{figure}[t]
\vspace*{-2.5cm}

\centerline{\scalebox{0.55}{\includegraphics{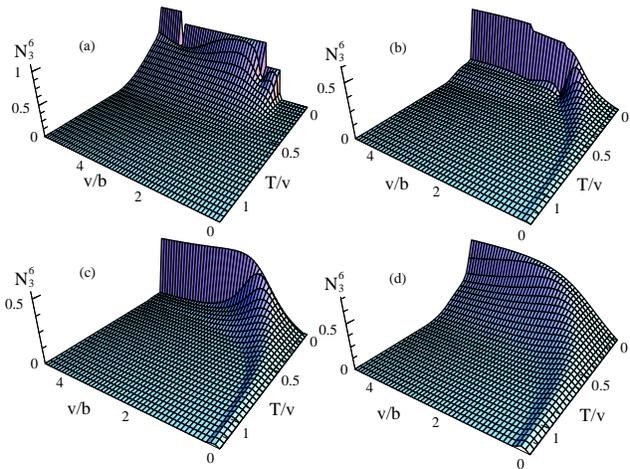}}}
    \vspace*{-8.5cm}

\caption{The most persistent global negativity as a function of temperature and
inverse magnetic field for $n=6$ and the full range interaction I, with
$v_-/v_+=0$ (a), 1/2 (b), 1 (c) and $\infty$ (d).}\label{f3}
\end{figure}

Results for the $XX$ case $v_x=v_y$ ($v_-=0$) are shown in fig.\ \ref{f2}. Here
the eigenstates of $H$ have definite values $M$ of $S_z$ and first order ground
state transitions $|M|\rightarrow |M|+1$ arise as $|b|$ increases (see fig.\
\ref{f3} and the ensuing description). For $\eta=2$ the ground state has
$|M|=1$ for $n=4$ and $n=6$, and is entangled and non-degenerate. The ensuing
thermal behavior is, accordingly, roughly similar to the previous case (results
for $n=3$, not shown, are very similar to those of fig.\ \ref{f1}). However,
for $n=4$ and $6$ we observe a more rapid initial decrease of {\it global}
negativities with increasing temperature, particularly noticeable for $n=4$ in
case II. This is due to the presence of a low lying entangled first excited
state, which becomes then mixed with the ground state already at low $T<T_c$,
thus reducing the negativity. Note that in contrast, the thermal average of the
interaction $\langle V\rangle$ {\it increases} initially with temperature for
$n=4$, since in this case this state has $M=0$ and is more correlated than the
ground state. For $n=6$ this state has instead $|M|=2$ and weaker correlations,
so that $\langle V\rangle$ also exhibits here an initial decrease. Another
novel aspect is the appearance of a small non-zero negativity for the first
non-adjacent pair density (qubits $ac$) in case II (right panels), which
vanishes at a low $T$. All other limits temperatures are slightly lower than
those of fig.\ \ref{f1}, although the ordering of those depicted remains
unchanged.

Fig.\ \ref{f3} offers a global view of the behavior with temperature and
magnetic field of the most persistent negativity for $n=6$ ($N^6_3$), for
increasing anisotropies in a type I interaction. For $v_x=v_y$ (a), global
negativities exhibit a {\it stepwise} increase at $T=0$ as $\eta$ increases,
reflecting the ground state transitions $|M|\rightarrow |M|-1$. The exact
energies are in this case given by (Eq.\ (\ref{HI}))
 \[ E_{SM}=bM-v[S(S+1)-M^2]/(n-1)+E_0\,,\]
so that for $v>0$ the ground state corresponds to $S=n/2$ and $|M|$ determined
by the ratio $\eta=v/|b|$. For $b>0$, a total of $[n/2]$ transitions
$M\rightarrow M+1$ occur therefore at
\[\eta=(n-1)/(2|M|-1)\,,\]
where $E_{SM}=E_{S,M+1}$, with $M=-n/2$ for $\eta<1$ (aligned state). The first
transition at $\eta=1$ marks then the appearance of entanglement at $T=0$ (with
$N^n_k=\sqrt{k(n-k)}/n$ for $M=-n/2+1$) and coincides with the onset of the
symmetry-breaking mean field solution. For $n=6$, the transitions occur at
$\eta=1,\frac{5}{3},5$.

For $T\rightarrow 0$, entanglement starts then only for $\eta>1$ and exhibits
drops at the critical ratios due to the degeneracy of the ground state. At
fixed low $T>0$, global negativities display accordingly smooth minima around
these values. Limit temperatures of global negativities are however independent
of $b$ in this case, so that for $\eta<1$ the system becomes entangled only for
$T>0$, as in the two-qubit case \cite{Ar.01,CR.04}, although negativities are
very small.

For small but non-zero anisotropies $v_-/v_+$, ground state transitions persist
(leading to discontinuities in $\langle S_z\rangle$ and the negativities), but
the ground state is no longer constant between transitions. The associated
negativities begin to vary smoothly between them and start already for $v>0$,
increasing linearly with $\eta$ for $\eta\ll 1$, following Eq.\ (\ref{nk1}). At
the same time, the limit temperature is no longer constant and the negativity
becomes non-zero for $\eta\rightarrow 0$ at any $T\ll b$. The concomitant
behavior can be appreciated in panel (b) for $v_-/v_+=\half$ ($v_y=v_x/3$)
where the ground state transitions occur at $\eta\approx 1.73$, $2.88$ and
$8.66$. For small $\eta$ we observe the tail corresponding to the entanglement
of the perturbed aligned state, while above the first transition the negativity
undergoes an abrupt initial decrease as $T$ increases from 0 due to the almost
degeneracy of the ground state. Note also that at low $T>0$, $N^6_3$ displays a
deep minimum in the vicinity of the first transition as $\eta$ increases.

For $v_-/v_+=1$ ($v_y=0$), the ground state transitions disappear, although the
energies of the ground and first excited states become almost degenerate for
large $\eta$ (where they approach the states $(|S,M_x\rangle\pm
|S,-M_x\rangle)/\sqrt{2}$ with $M_x=S=n/2$ and $|S,M_x\rangle$ the eigenstates
of $S^2$ and $S_x$), with an energy splitting $\Delta E\propto v\eta^{-n}$ in
this limit. This leads again to a rapid initial drop of the negativity as $T$
increases from $0$. The final effect for low $T>0$ and $\eta\agt 1$ is a {\it
decrease} of the negativity for increasing $\eta$. Finally, the behavior in the
fully anisotropic case $v_-/v_+=\infty$ ($v_y=-v_x$) is completely smooth as
either $T$ or $\eta$ increases. All global negativities increase as $\eta$
increases at $T=0$ and decrease smoothly as $T$ increases.

\begin{figure}[t]
\vspace*{-2.5cm}

 \centerline{\scalebox{0.55}{\includegraphics{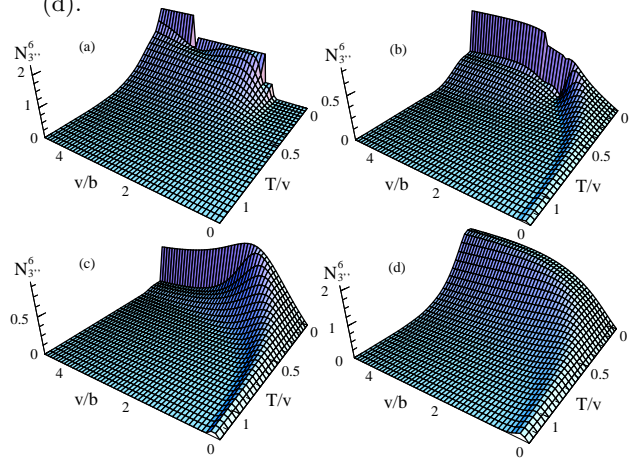}}}
    \vspace*{-8.5cm}

\caption{Same quantities and details as in fig.\ \ref{f3} for the nearest
neighbor interaction II.}\label{f4}
\end{figure}

The corresponding results for the type II interaction (fig.\ \ref{f4}) exhibit
the same behavior. The most persistent negativity in all cases depicted is that
of the non-adjacent 3+3 partition $N^6_{3''}=N_{ace-bdf}$, which for
$\eta\rightarrow 0$ and $T\ll b$ increases for non-zero anisotropy as in Eq.\
(\ref{nk2}) with $r_k=4$ (for the other partitions
$N^6_{2'},N^6_{2''},N^6_{3'}$ we have $r_k=1+\sqrt{3},2\sqrt{2},1+\sqrt{5}$
respectively). In the $XX$ case (panel a), the ground state exhibits again
three transitions $M\rightarrow M+1$ at $\eta=1$, $1/(\sqrt{3}-1)\approx 1.37$
and $1/(2-\sqrt{3})\approx 3.73$, corresponding to energies $E_{-3}=-3b$,
$E_{-2}=-2b-v$, $E_{-1}=-b-\sqrt{3}v$ and $E_0=-2v$, where $E_M$ denotes the
lowest energy for a given $M$. The first transition occurs again at the same
value $\eta=1$ $\forall$ $n$, and the ground state for $\eta<1$ is again the
aligned state, so that the $T=0$ negativity starts only for $\eta>1$ and
increases stepwise. For $v_-/v_+=1/2$ (b) the transitions occur at $\eta\approx
1.73$, 2.31 and 6.28. The negativity exhibits in this case a non-monotonous
behavior at $T=0$, showing a maximum and a minimum just before the first
transition, the latter becoming very pronounced as $T$ increases. The same
previous effects are seen for $v_-/v_+=1$ (c), where the energy gap $\Delta E$
between the ground and first excited states decreases again as $v\eta^{-n}$ for
$\eta\gg 1$, while for $v_y=-v_x$ (d) the behavior is again completely smooth.

\begin{figure}[t]
\vspace*{-1.5cm}

 \centerline{\scalebox{0.55}{\includegraphics{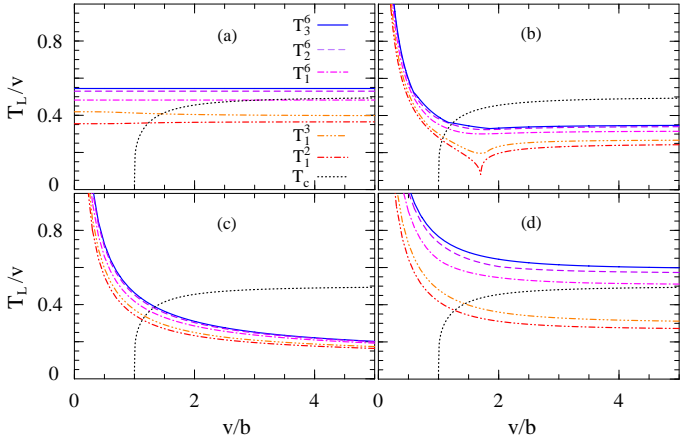}}}
    \vspace*{-9.5cm}

\caption{Limit temperatures $T_L\equiv T^m_k$ for non-zero negativities
$N^m_k$, for the full range interaction I with $v_-/v_+=0$ (a), $1/2$ (b), $1$
(c) and $\infty$ (d). The dotted line depicts the mean field critical
temperature $T_c$.}\label{f5}
\end{figure}

Beneath the surfaces of figs.\ 3-4 lie those of the remaining global and
reduced negativities. We plot in fig.\ \ref{f5} the corresponding main limit
temperatures for type I interaction. As previously shown, in the $XX$ case
(panels a) $T^n_k/v$ is constant for all global negativities. This is also
approximately true for limit temperatures of all reduced negativities in I.
However, for non-zero anisotropies (panels b,c,d), {\it all} limit temperatures
in I diverge for $\eta\rightarrow 0$, in agreement with our previous
discussion. This is in marked contrast with the behavior of the mean field
critical temperature, which vanishes for $\eta<1$. Limit temperatures are not,
however, proportional to $|b|$ in this limit and ratios $T^n_k/|b|$ actually
vanish for $\eta\rightarrow 0$.

Whereas in panels (a) and (d) limit temperatures of global negativities become
close to $T_c$ for large $\eta$, in (b) and particularly c) they become
substantially {\it lower} than $T_c$,  due to the quasi-degeneracy of the two
lowest energy levels in this region. This indicates that symmetry-breaking is
not necessarily an indication of entanglement at finite temperature. Besides,
limit temperatures {\it are not necessarily smooth functions of} $\eta$, as
occurs for instance in case (b), where $T^6_3$ exhibits two slope
discontinuities at $\eta\approx 1.17$ and 1.83, being minimum at the last
value. These transitions reflect the changes in the lowest eigenvalue of
$\rho^{t_p}$, arising from level crossings, and become more noticeable in the
limit temperatures of reduced negativities, as seen for $T^3_1$ and
particularly $T^2_1$, which exhibits a deep minimum at $\eta\approx 1.7$. The
thermal behavior of the associated negativity in the vicinity of these
crossovers can be more complex than in figs.\ 1-2, and may exhibit a deep
minimum followed by a maximum before vanishing, which may even evolve into a
complete vanishing plus a reentry.

\begin{figure}[t]
\vspace*{-1.5cm}

 \centerline{\scalebox{0.55}{\includegraphics{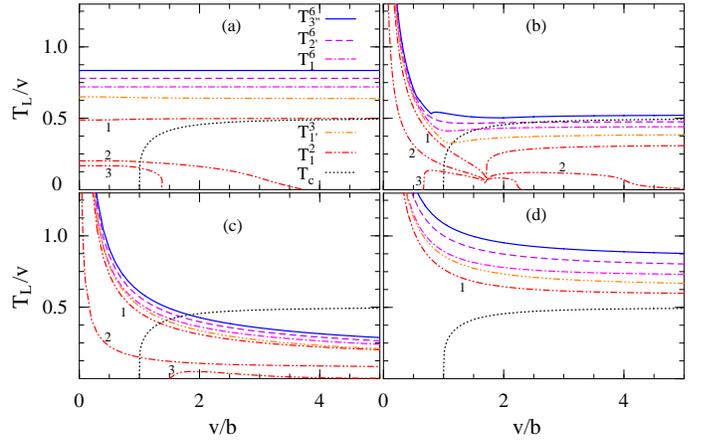}}}
    \vspace*{-9.5cm}

\caption{Same quantities and details as in fig.\ \ref{f5} for the nearest
neighbor interaction II. Labels 1,2,3 indicate results for the adjacent and the
two first non-adjacent pair densities respectively.}\label{f6}
\end{figure}

Results for type II interaction (fig.\ \ref{f6}) are quite similar. Limit
temperatures of global negativities and of reduced densities of adjacent qubits
are higher than in I for the same value of $v$, although they exhibit the same
behavior and still fall well below $T_c$ in (c). With the exception of case
(d), reduced pair densities of {\it non-adjacent} qubits (i.e., $(a,c)$ (2) and
$(a,d)$ (3)) have as well non-zero negativities  at least for some field
intervals, although they possess low limit temperatures that depend strongly on
the magnetic field even in (a). For $n=6$, qubits $(a,c)$ in (a) are entangled
at $T=0$ just between the first and third transition ($1<\eta<3.73$) while
qubits $(a,d)$ just between the first and second transition ($1<\eta<1.37$),
although for $T>0$ both become weakly entangled also for $\eta<1$ (reentry
effect). Case b) is more complex since it exhibits crossovers as in type I,
reflected in the appearance of minima and slope discontinuities in the limit
temperatures. The highest one $T^6_{3''}$ has minima with slope discontinuities
at $\eta\approx 0.8$ and 1.99, the latter being the absolute minimum. That of
the reduced density of adjacent qubits displays actually a discontinuity at the
minimum which is the signature of a {\it reentry effect} (for $1.71\alt\eta\alt
1.72$, $N^2_1$ exhibits a small reentry as $T$ increases after its first zero,
originating a discontinuity in the final limit temperature at $\eta\approx
1.71$). Those of non-adjacent qubits $ac$ (2) and $ad$ (3) exhibit the same
behavior in the vicinity of $\eta=1.7$. That of qubits $ac$ becomes infinite
for $\eta\rightarrow 0$, like $T^2_1$, but vanishes after the third $T=0$
transition ($\eta\agt 6.28$), while that of $ad$ is non-zero just for
$0.68\alt\eta\alt 2.31$ (below the second $T=0$ transition). In c) qubits $ac$
become entangled $\forall$ $\eta>0$, and their limit temperature, though lower,
exhibits the same behavior as that of adjacent qubits, while qubits $ad$ become
entangled just for $\eta\agt 1.5$ up to a very low temperature. Finally, in d)
non-adjacent pairs become separable, while the other limit temperatures become
all higher than $T_c$.

\section{Conclusions}
In this work we have investigated the thermal entanglement of $n$-spin systems
by evaluating the set of negativities associated with all possible bipartite
splittings of the system and subsystems. Entanglement is then seen to decay for
increasing $T$ through a cascade of limit temperatures that determine the onset
of separability of the different bipartitions. For the cases here considered,
the most persistent global negativity is that of the most symmetric bipartition
($N^n_{[n/2]}$) in the fully connected case I, and the same occurs in the
nearest neighbor case II provided {\it non-contiguous} qubits are chosen in
each partition. Negativities of reduced densities and of weakly interacting
splittings vanish obviously earlier, so that there is always some final
interval where only {\it global} entanglement survives. The behavior with
temperature and magnetic field of the most persistent negativity is rather
similar in I and II for the cases considered after adequate scaling of coupling
strengths ($v_{ij}\propto v/n$ in I and $\propto v$ in II) but depends strongly
on the anisotropy.

In all $XXZ$ models, we have shown that limit temperatures of global
negativities are {\it strictly independent} of the (uniform) applied magnetic
field $b$, for {\it any} value of the total qubit number $n$, even though the
negativity may exhibit a stepwise variation with the field at $T=0$. This
implies in particular that one cannot expect an agreement for all fields
between these limit temperatures and the mean field critical temperature $T_c$,
even for {\it large $n$}, as the latter always {\it vanishes} for sufficiently
large fields. The lack of agreement persists in anisotropic models, where limit
temperatures for global entanglement were shown to become large for large
fields, even though the negativity tends to zero  in this limit. In this case
we have shown explicitly that for large fields ($\eta<1$) and sufficiently low
$T$, negativities of most symmetric partitions remain finite even for large $n$
(Eqs.\ (\ref{nk1})-(\ref{nk2})). Nevertheless, for $\eta>1$ the negativity is
normally seen to become relatively small above $T_c$ (figs.\ 1-2), so that in
this sense an approximate agreement with the mean field picture is recovered.
Negativities may also vanish for $T<T_c$ when ground state degeneracies (exact
or approximate) are present, as seen in cases b and c in figs. 5-6.
Finally, it is to be remarked that limit temperatures may exhibit slope or full
discontinuities for increasing fields for finite anisotropies, reflecting
crossovers between different entanglement regimes, which may become more
pronounced for those of reduced negativities. The present study provides
therefore a more complete understanding of the way finite spin systems loose
their quantum correlations due to standard (i.e., Boltzmann like) thermal
randomness. Other aspects of the problem are presently under investigation.

{\it Acknowledgments}. RR and NC acknowledge support, respectively, from CIC
and CONICET of Argentina.

\end{document}